\documentstyle[emul_rec,apjfonts,epsf]{article}
\newcommand{\etal}{et al.~}
\begin{document}

\title{An eclipsing millisecond pulsar with a possible main-sequence
companion in NGC~6397}

\author{N.~D'Amico,\altaffilmark{1,2}
A.~Possenti,\altaffilmark{1}
R.~N.~Manchester,\altaffilmark{3} 
J.~Sarkissian,\altaffilmark{4}  
A.~G.~Lyne\altaffilmark{5}
and F.~Camilo\altaffilmark{6}
}
\medskip

\affil{\altaffilmark{1}Osservatorio Astronomico di Bologna,
Via Ranzani 1, 40127 Bologna, Italy}
\affil{\altaffilmark{2}Istituto di Radioastronomia del CNR,
Via Gobetti 101, 40126 Bologna, Italy}
\affil{\altaffilmark{3}Australia Telescope National Facility,
CSIRO, PO Box 76, Epping, NSW 2121, Australia}
\affil{\altaffilmark{4}Australia Telescope National Facility,
CSIRO, Parkes Observatory, PO Box 276, Parkes, NSW 2870, Australia}
\affil{\altaffilmark{5}University of Manchester, Jodrell Bank
Observatory, Macclesfield, Cheshire SK11~9DL, UK}
\affil{\altaffilmark{6}Columbia Astrophysics Laboratory, 
Columbia University, 550 West 120th Street, New York, NY 10027}

\bigskip

\begin{abstract}

We present the results of one year of pulse timing observations of
PSR~J1740$-$5340, an eclipsing millisecond pulsar located in the globular
cluster NGC~6397.  We have obtained detailed orbital parameters and a
precise position for the pulsar. The radio pulsar signal shows frequent
interactions with the atmosphere of the companion, and suffers significant
and strongly variable delays and intensity variations over a wide range of
orbital phases.  These characteristics and the binary parameters indicate
that the companion may be a bloated main-sequence star or the
remnant (still filling its Roche lobe) of the star that spun up the
pulsar. In both cases, this would be the
first binary millisecond pulsar system with such a companion.

\end{abstract}

\keywords{globular clusters: individual (NGC~6397) --- pulsars:
individual (PSR~J1740$-$5340) --- binaries: close}

\section{Introduction}

The millisecond pulsar J1740$-$5340 in the globular cluster NGC~6397
(\cite{dlm+01}) is a member of a binary system with a relatively wide
orbit of period 1.35 days, a companion with mass $M_c >0.19$\,M$_\odot$, and is
eclipsed for about 40\% of its orbit at a frequency of $\nu = 1.4$\,GHz.  In
much tighter eclipsing binaries, like those containing PSRs~B1957+20
(\cite{fbb+90}), B1744$-$24A (\cite{lmd+90,nt92}) and J2051$-$0827
(\cite{sbl+96}), all having companions of mass $M_c<0.10$\,M$_\odot$, 
the eclipses are believed to be caused by a wind resulting from ablation 
of the companion by a relativistic pulsar wind. PSR~J1740$-$5340 is in 
a wide orbit and the wind energy density impinging on the companion 
is significantly less than that estimated for the close eclipsing systems, 
and is unlikely to drive a wind of sufficient density off a degenerate 
companion. In the discovery paper, D'Amico et al. (2001) \nocite{dlm+01} 
argued that the companion could be an unevolved star of mass comparable 
to the turnoff mass of the cluster, $\sim 0.8$ M$_{\odot}$, 
releasing a wind sufficiently dense to produce the observed eclipses. 
Such a hypothesis, which would make PSR~J1740$-$5340 an
unusual system, can be ultimately tested with the optical
identification of the pulsar companion.

In this Letter we report on the first year of pulse timing observations
of this pulsar. We provide timing parameters, including the position, and
we report on further observational evidence of the interaction of the
pulsar signal with the companion atmosphere.

\section{Observations and Results}

Regular timing observations have been made since 2000 July with the 
Parkes radio telescope, using the center beam of the multibeam receiver
at 1.4 GHz. The observing system is the one used in the discovery 
observations (\cite{dlm+01}). The effects of interstellar dispersion 
are minimized with a filter bank having 512 $\times$ 0.5 MHz frequency 
channels for each of two linear polarizations.
After detection, the signals from individual channels are added in
polarization pairs, integrated, 1 bit-digitized every 125 $\mu$s, and then
recorded on magnetic tape. Pulse times of arrival (TOAs) are determined
offline by fitting a standard pulse profile having a high signal-to-noise
ratio to the observed profiles and analyzed using the program {\sc
tempo}\footnote{http://pulsar.princeton.edu/tempo.}  and the DE200 solar
system ephemeris (\cite{sta82}). A simple Keplerian model of a binary system
was used in the analysis.

\begin{figure*}
\plotfiddle{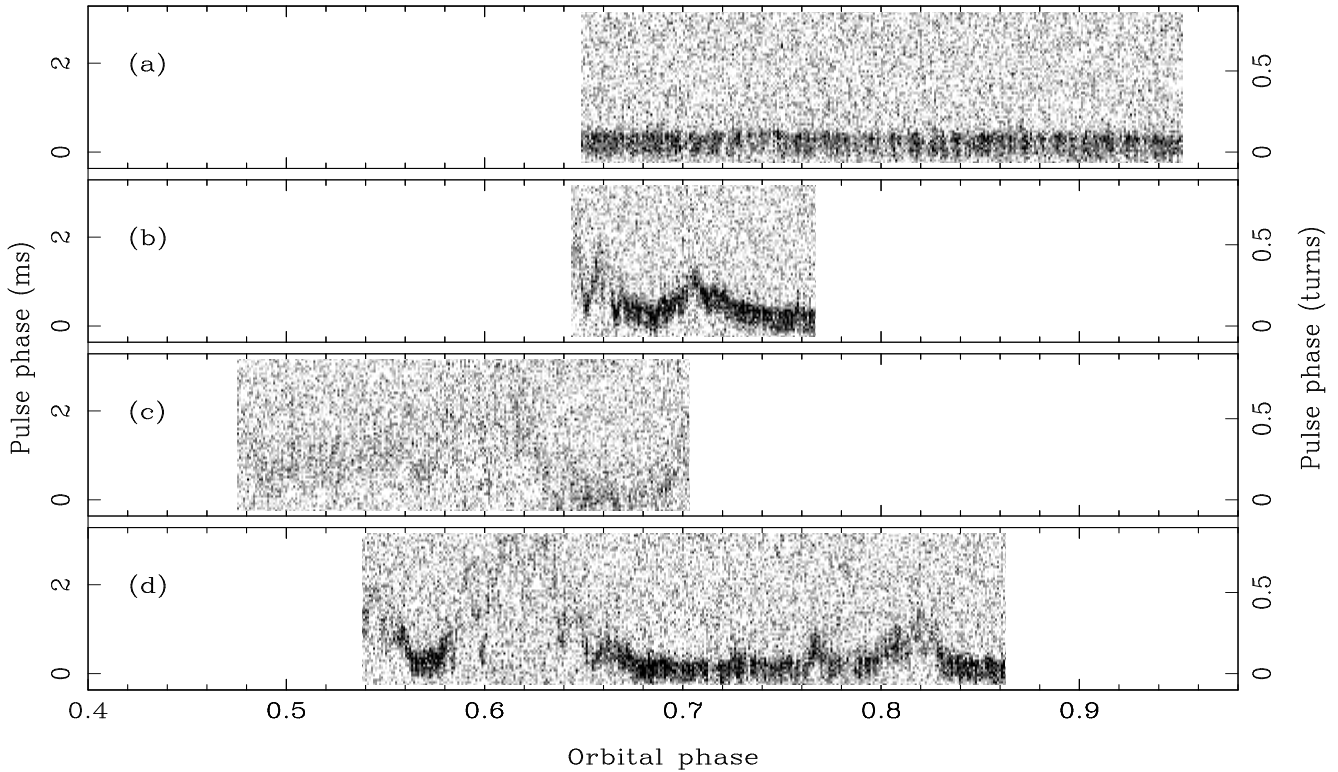}{9.5truecm}{0}{93}{80}{-340}{-270}
\figcaption[f1_rot.eps]{\footnotesize{
Observed signal intensity at 1.4 GHz as a function of
orbital phase and pulsar phase for four long observations away from the
nominal eclipse region. 
The data are processed in contiguous integrations of 120 s
duration. (a) $\sim 10$ hr observation starting on 2000 December 3 at 22:23
UT; (b) $\sim 4$ hr observation starting on 2001 March 1 at 22:32 UT; (c)
$\sim 7$ hr observation starting on 2001 March 5 at 18:31 UT; (d) $\sim 11$
hr observation starting on 2001 March 12 at 15:02 UT.}}
\end{figure*}

For about six months we routinely collected data on this pulsar several
times a week using an integration time of 30 min for each observation. From
the orbital parameters obtained during the confirming observations in 2000
July, we knew that the pulsar was eclipsed in the orbital phase interval
0.05--0.45, so we have generally made observations at epochs when the
predicted
orbital phase was outside this interval. In this way we have obtained a
reasonable number of good quality integrated pulse profiles (with
signal-to-noise ratios in the range 10--25 and typical timing errors in the
range 10--40 $\mu$s), while some low quality profiles were rejected.  Most
of the latter were obtained at epochs relatively close to the predicted
eclipse edge, but some were taken far from the nominal eclipse region, at
orbital phases where the pulsar was expected to be detected clearly. We also
noted that, while we could obtain a coherent timing solution over several
months, the resulting post-fit rms scatter of the residuals was large
compared to the typical nominal timing uncertainties of individual TOAs,
suggesting the presence of systematic effects.

In order to probe the hypothesis that this behavior could be due to
the interaction of the pulsar radiation with the companion
atmosphere, we have recently initiated a series of long
observations starting close to the eclipse edge near the descending
node, and lasting for a substantial fraction of the orbit in
the non-eclipsing region.

The orbital period is $\sim$ 32 hr and the pulsar is visible at Parkes 
for $\sim$ 11 hr on a given day. Thus far we have obtained four relatively
long observations.  The results are shown in Figure~1. In one case 
(Fig.~1a) the pulsar signal was relatively stable in intensity and 
pulse phase, while in the other three cases (Figs.~1b, c and d) 
significant excess propagation delays of up to $\sim$ 3 ms and 
strong intensity variations are observed.
In particular, in the case of Figure~1c, the pulsar signal was always rather
weak and the pulse was about two times wider than usual, covering
about 30$-$50\% of the rotational phase. The correlation between
delay peaks and flux reduction suggests that interstellar scintillation
is not responsible for this behavior. These fluctuations are
comparable to those observed at a lower frequency ($\sim$ 400 MHz) in
PSR~B1957+20 near the eclipse edges around orbital phases 0.2 and 0.3
(\cite{fbb+90}). In PSR~J2051$-$0827 (\cite{sbl+96}) flux density variations 
and variable propagation delays are seen at 1.4 GHz when the pulsar 
signal is detectable through the eclipse region. PSR~B1744$-$24A 
displays similar irregularities also far from the 
usual eclipse region, with typical delays of 
$\sim 300 \mu$s (\cite{nt92}). Instead, much larger excess delays, 
$\ga$ 1 ms, are observed at nearly all orbital phases in the case
of PSR~J1740$-$5340. These observations suggest that the pulsar 
is in fact orbiting within an extended envelope of matter released 
from the companion.

We investigated the frequency-dependent behavior of the
observed delay events by splitting the 256-MHz bandwidth available at
1.4 GHz into two adjacent subbands.  As can be seen from Figure~1,
significant pulse broadening and signal-to-noise reduction is usually
associated with large delays.  We have therefore limited our analysis to
small delays only.  Figure~2 shows an example of the measured excess
delays in the two subbands.  The small panel also shows these
measurements plotted against each other.  The best-fitting straight
line corresponds to $\Delta t \propto \nu^{-2.02\pm 0.30}$
for the excess delays, consistent with the $\nu^{-2}$ dependency
expected if the delays are due to dispersion in an ionized medium. If
entirely due to dispersion, the excess propagation delays of up to
$\sim 3~{\rm ms}$ visible in Figure~1 would correspond to electron
column density variations of $\sim 1.5\times 10^{18}~\Delta t_{-3}$
cm$^{-2}$, or $\Delta{\rm DM}\sim 0.47~\Delta t_{-3}$ cm$^{-3}$pc
($\Delta t_{-3}$ being the delay at 1.4 GHz in ms).  This corresponds
to a pulse broadening of $0.10~P~\Delta t_{-3}=0.36\Delta~t_{-3}$ ms 
over the receiver bandwidth (to be compared with an intrinsic 
pulse FWHM of $0.17~P$) and may be responsible for most 
of the signal attenuation and pulse broadening observed near the delay peaks.

From Figure~1 we can also see that far from the eclipse region 
the typical time scale of the rise to maximum delay is of 
order 1$-$2 hr. This indicates that observations of $\ga$ 3 hr 
are required in order to identify reliably a 
non-delayed portion of an observation and to obtain a bona fide
TOA. This explains why a timing solution mainly based on relatively
short integrations may be affected by systematic effects.   

In order to obtain a reliable phase-connected timing solution for this
pulsar we have selected by visual inspection among all the data 
collected at 1.4 GHz in the last 12 months only those TOAs 
with no indication of delay and those corresponding to profiles 
not affected by pulse broadening.  
In this way we have obtained a solution with
a post-fit rms residual of $\sim$ 120 $\mu$s. The corresponding model
parameters are reported in Table~1. Figure~3a shows the residuals for
all TOAs collected whereas Figure~3b shows those for the selected TOAs
only, all relative to the model in Table~1.
The dispersion measure (DM) was fitted using TOAs from two
adjacent subbands from the stable long integration shown in Figure~1a.
A fit for orbital eccentricity may be biased because
only $\sim 60\%$ of the orbit is sampled. Nevertheless a formal fit
results in an upper limit of $e<10^{-4}$ ($3\sigma$), and in deriving
the parameters listed in Table~1 we have fixed $e=0$. Further observations
may better constrain this limit. Similarly, a longer data span will 
be needed for investigating possible changes in the orbital parameters, as 
seen e.g. in PSR~B1957+20 (Nice, Arzoumanian \& Thorsett 2000) \nocite{nat00}.

\section{Discussion}

PSR~J1740$-$5340 has the longest orbital period ($32$ hr) and the
most massive minimum companion mass ($0.19$ M$_\odot$, assuming
a neutron star mass of 1.40 M$_\odot$) among the 10
known eclipsing pulsars (Nice, Arzoumanian \& Thorsett 2000\nocite{nat00};
Camilo \etal 2000\nocite{clf+00}). 
When combined with the typical duration of
the eclipse at 1.4 GHz ($\sim 13$ hr), these parameters imply an
eclipse radius $R_E$ larger than the orbital separation
$a$ for any value of the orbital inclination $i$. $R_E$ is defined
as the chord subtended on the orbital circumference of radius $a$ 
by the angle between the orbital phase of eclipse ingress (or egress) 
and the orbital phase 0.25. Table~2 shows the
relevant parameters associated with different companion types: a 0.19
M$_{\odot}$ He-white dwarf (WD) (corresponding to $i=90\arcdeg$) 
and a range of main sequence (MS) stars spanning
the mass interval $M_{ms} \sim 0.19-0.8$ M$_{\odot}$ 
(the latter value being the maximum compatible with the age of NGC~6397).  
As can be seen, $R_E\sim 7.5$ R$_\odot$ $>$ $a \sim 6.5$ R$_\odot$.  

\medskip
{\plotone{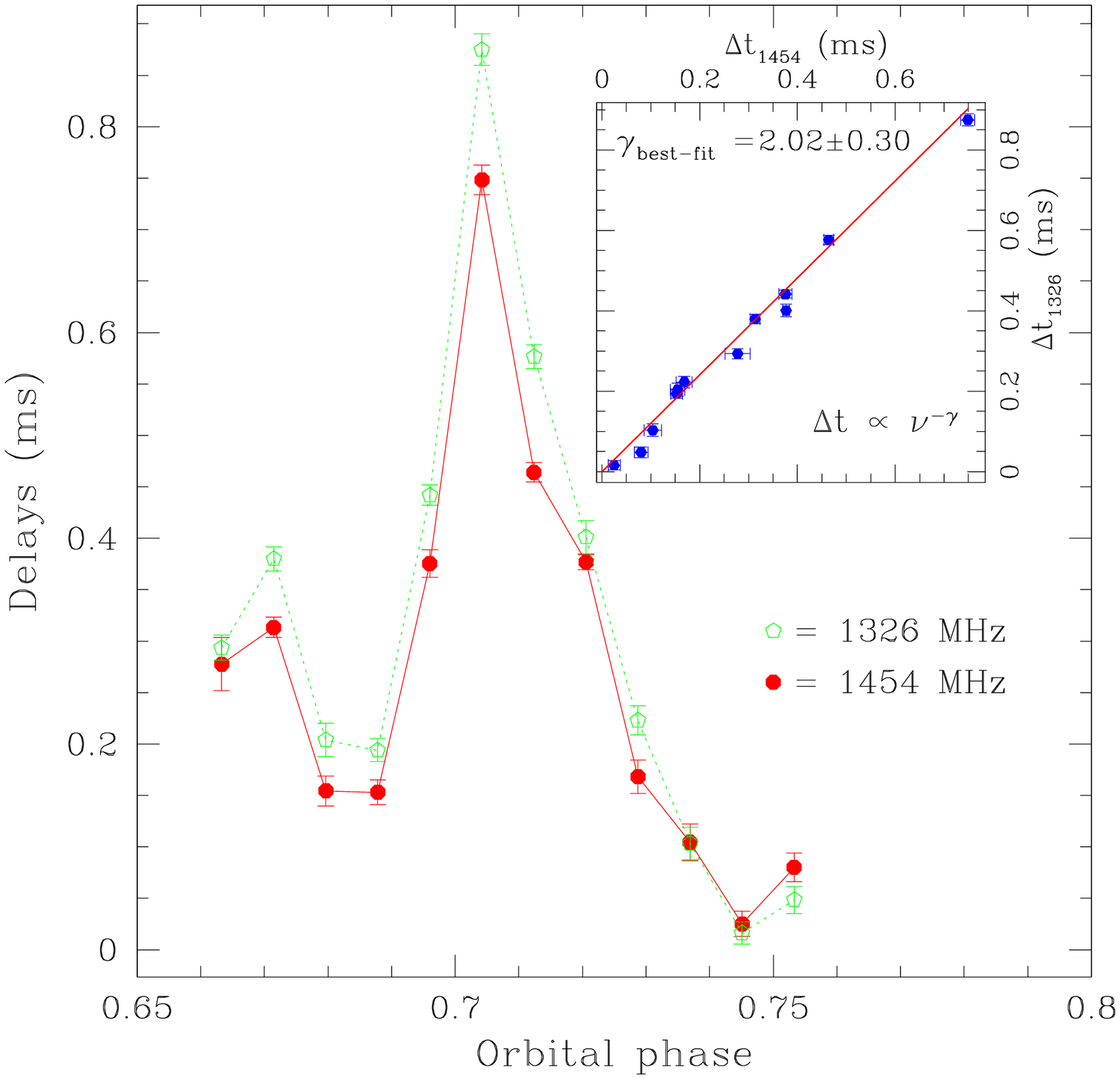}
\figcaption[f2-color.eps]{\footnotesize{
Excess group delays measured in two 128 MHz-wide bands 
centered at 1454 MHz ({\it filled symbols\/}, connected by a {\it solid
line\/}) and at 1326 MHz ({\it open symbols\/}, connected by a 
{\it dotted line\/}) for an event occurring at orbital phases 
0.65$-$0.75 on 2001 March 1 (see Fig. 1b).
The delays in the two subbands are fitted 
({\it inserted panel\/}) with a straight line of slope 1.21$\pm 0.04$, 
corresponding to a frequency dependence for the delays 
($\Delta t \propto \nu^{-\gamma}$) with a best-fit power-law 
index $\gamma =2.02\pm 0.30.$}}
\skip 1.5truecm}
\medskip

From the observation in
Figure~1a, we infer that there are no systematic electron column
density variations over a wide orbital phase interval.  Hence the neutron
star is not orbiting inside a steady corona of ionized matter
surrounding the companion star. More likely, the pulsar is just
skimming a large envelope of matter, whose clumpness and varying
shape at distances from the companion comparable with $R_E$ could be
triggered by the movement and the energy flux of the pulsar itself.
The irregular features appearing in the three observations of 2001 March 
(Figs.~1b--1d) could for example represent this kind
of interaction.

The values of the Roche lobe radius $R_L$ span the interval
1.3--2.2\,R$_\odot$ (\cite{egl83}).  Because $R_E > a > R_L$, the
matter causing the eclipses escapes the gravitational influence of the
companion star and therefore must be continuously replenished.
The large position offset of PSR~J1740$-$5340 with respect to the 
center of NGC~6397 ($r\sim 0\farcm55$) limits the unknown contribution 
to the observed period derivative $\dot{P}$ due to acceleration in
the globular cluster potential. Assuming a King model
(1962\nocite{k62}) for the central mass density, core radius $r_c=0\farcm05$
and central luminosity density (in units of ${\rm L_\odot}$/pc$^3$) 
$\log(\rho_L[0])=5.6$ (\cite{h96}), 
distance $d=2.6~{\rm kpc}$ (Reid \& Gizis 1998\nocite{rg98}), and 
a mass-to-luminosity ratio $M/L=3.3$ (\cite{pm93}), we derive (see
e.g. Camilo et al. 2000\nocite{clf+00})
$|\dot{P}_{\rm acc}|<4\pi G/(9c)(r_c/r)r_cd(M/L)\rho_L(0)P\sim 10^{-20}$.
Even accounting for the uncertainties in some of the above parameters 
(e.g., $r_c$), the observed $\dot{P}$ (Table~1) should be only slightly
affected
by $\dot{P}_{\rm acc}$. Thus we can use $\dot{P}$ to infer the rotational
energy 
loss of the pulsar, $\dot E\sim 1.4\times 10^{35}$ ergs s$^{-1}$.
A He-white dwarf of minimum mass
$M_{wd}=0.19~{\rm M_\odot}$ with an effective temperature $T_{wd}=10^4$K
has radius $R_{wd}=0.033~{\rm R_\odot}$ (\cite{dsbh98,sar+01}).
Because the radius decreases with increasing temperature, we use $R_{wd}$ as
an upper limit. At the distance of $6.0~{\rm R_\odot}$, the power impinging
on the degenerate companion would be $W_{wd} \la 10^{30}$ ergs s$^{-1}$. 
This would allow release of an ionized wind from the
degenerate companion at a maximum rate $\dot
M_{wd,max}=W_{wd}(R_{wd}/GM_{wd})\la 1.6\times 10^{-12}\; {\rm
M_\odot y^{-1}}.$ 

\medskip
{\plotone{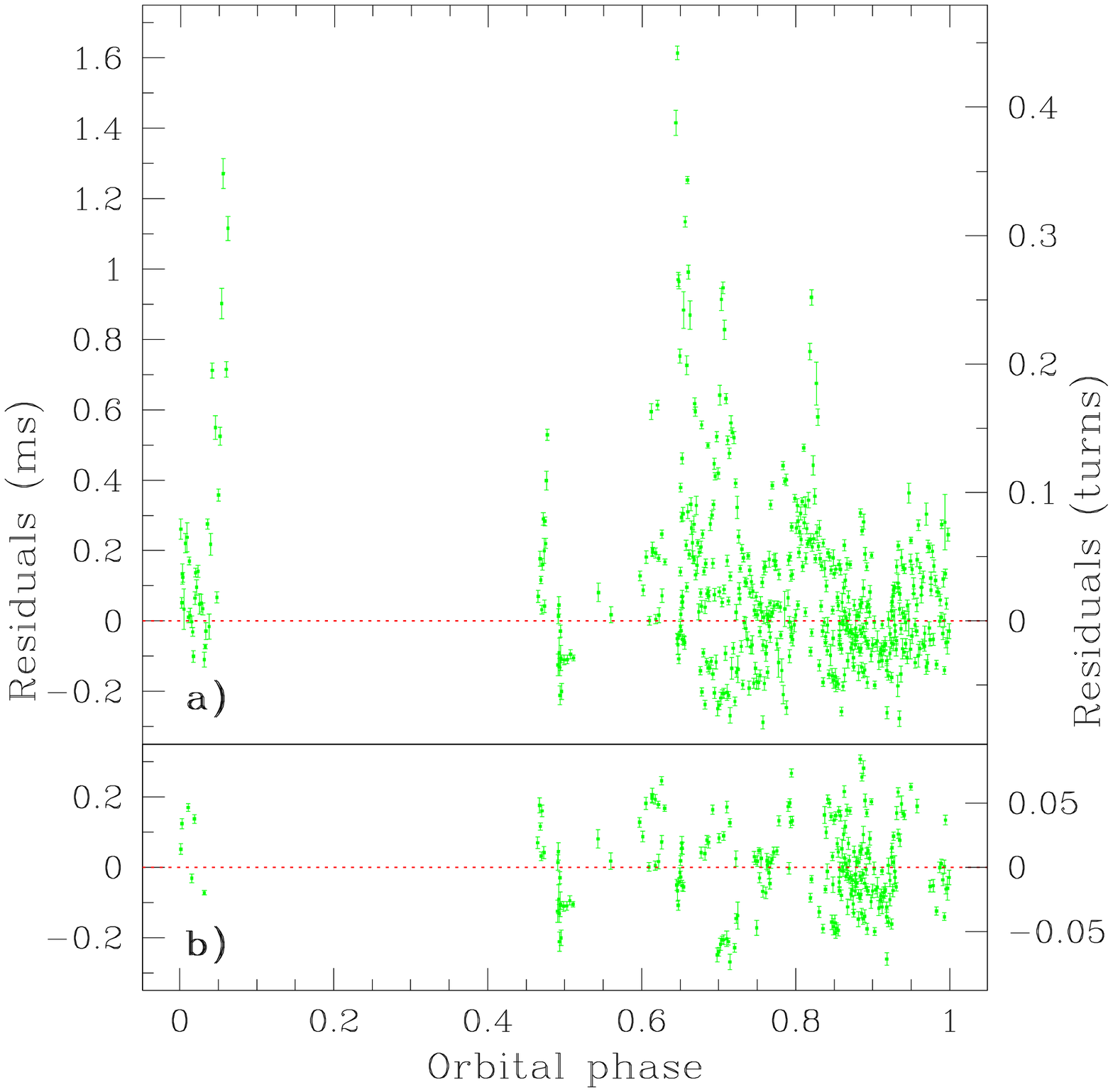}
\figcaption[f3-color.eps]{\footnotesize{
(a) Timing residuals of all the TOAs collected (525 points 
derived from 46 observations of length ranging from $\sim$ 30 min up to
$\sim$ 11 hr) relative to the timing model in Table~1.
(b) Timing residuals of the 241 TOAs  
used to derive the phase-connected solution reported in Table~1. 
The scales of the two panels are identical.
The post-fit timing residuals in panel b, while smaller on average than
those in panel a, are still noticeably large compared to the nominal TOA
uncertainties. This effect may be caused by small-scale electron density 
fluctuations not modeled in our present fits.}}
\skip 1.5truecm}
\medskip

Assuming isotropic emission, mass continuity
implies that the companion star releases mass at a rate
$\dot M_{c}=4\pi R_E^2\rho (R_E)v_f.$ Using the
values in Table~2 and the density at the eclipse radius inferred from
the observation of the excess delays assuming completely ionized
matter ($\rho [R_E]\sim 3\times 10^{-18} \Delta t_{-3}$ g cm$^{-3}$),
we obtain $\dot M_{c}\sim 1.5\times 10^{-11}~\Delta t_{-3}~v_{f,8}~{\rm
M_\odot y^{-1}}$ where $v_{f,8}$ is the wind velocity at $R_E$ in units of
10$^8$ cm s$^{-1}$ (typical order of magnitude of the escape velocity
from the surface of the companion). Thus, only in the case of low
terminal velocity of the wind ($v_f \la 100$ km s$^{-1}$,
corresponding to improbable electron temperatures of $\la 200$ K at
$R_E$) could the pulsar flux sustain the ablation.  
More massive WDs would further exacerbate these difficulties.

The hypothesis of a MS companion appears more viable: such
a companion could have been acquired as a result of an exchange
interaction in the cluster core (Sigurdsson \& Phinney 1993\nocite{sp93}; 
Heggie, Hut \& McMillian 1996\nocite{hhm96}), 
a scenario possibly supported also by the
position of PSR~J1740$-$5340 $\sim 11$ core radii from the center of
NGC~6397. This object is the second farthest pulsar from a
cluster core (in units of core radii) among the 36 pulsars in globular 
clusters with precisely determined positions (after PSR~B2127+11C in 
M15, located at $\sim 13$ core radii).

The unperturbed radius $R_{ms}$ of a MS star of mass
$M_{ms} \sim 0.19$--$0.8~{\rm M_{\odot}}$
is $\sim 6$--30 times greater than the maximum possible
radius for a WD (Table~2). 
The energy required for transporting a unit mass from the
surface of the companion to its Roche lobe radius $R_{L,*}$ is
$U_{*}\sim GM_{*}(R_{*}^{-1}-R_{L,*}^{-1})$. Given this, the corresponding
maximum ablation rate for a MS companion is $\dot M_{ms,max}\sim
(R_{ms}/R_{wd})^2(U_{wd}/U_{ms}){\dot M}_{wd,max}\sim 200$--$6000{\dot
M}_{wd,max}=(3$--$90)\times 10^{-10}~{\rm M_\odot y^{-1}}$. Therefore, only
0.2--5\% of the impinging power would be needed to sustain
the inferred mass loss rate $\dot M_{c}$.
The bulk of the pulsar energy would thus be available for heating the
companion photosphere and perhaps inflating its radius up to Roche lobe, in
turn making the release of the wind even easier. 
This model should be more easily applicable to a light MS star, 
as large convective envelopes favor bloating of the star (\cite{p91}). 

Another hypothesis for explaining the large mass loss rate $\dot M_{c}$
is that the companion is an evolved star that spun up the millisecond
pulsar and that is presently undergoing the presumed final stages of mass loss.
In this case, mass accretion and spinning up of the pulsar would now be 
inhibited by the pulsar wind flux which could expel the matter overflowing 
from the Roche lobe of the companion (see for instance Burderi \etal 2001 and
references therein\nocite{bdm+01}). If substantiated, this would be the first
confirmed example of a recently born millisecond pulsar.  
The apparent rather young spin-down age of PSR~J1740$-$5340 
($\tau_c = P/2\dot P\sim 350$ Myr, assuming as mentioned earlier
that $\dot{P}_{\rm acc}$ is negligible), and its surface magnetic field
($B = 3.2\times 10^{19} [P \dot P]^{1/2} \sim 8\times 10^8$ G) 
would support this model. However, the
future evolution of the binary (companion evolving into a WD?) remains unclear.
Detailed simulations of the system are required for comparing this or other
more exotic scenarios (Burderi, D'Antona \& Burgay 2001, in preparation) 
with the parameters derived from radio and optical observations. 

Ferraro et al. (2001)\nocite{fe+01} report the possible identification of the
optical companion to PSR~J1740$-$5340 with a bright variable star and
Grindlay et al. (2001)\nocite{ghemc01} have detected with {\sl Chandra}
the X-ray counterpart of the millisecond pulsar.
The availability of data in a large spectral band, from
radio to optical and X-rays, should help understanding 
this exotic system and the mechanism(s) responsible for the radio eclipses.

\acknowledgements
\small{The Parkes radio telescope is part of the Australia Telescope which is
funded by the Commonwealth of Australia for operation as a National Facility.
We thank the staff at Parkes for their support of this project, and
L. Burderi for carefully reading the manuscript and for stimulating
discussions.
N.D'A. and A.P. are supported by the Ministero della Ricerca Scientifica
e Tecnologica (MURST). F.C. is supported by NASA grant NAG5-9095.}

\begin{deluxetable}{ll}
\scriptsize
\tablewidth{0.98\columnwidth}
\tablecaption{\label{tb:tpar} {Parameters of the PSR~J1740$-$5340 system}}
\tablecolumns{2}
\tablehead{\colhead{Parameter}&\colhead{Value}}
\startdata
Right Ascension (J2000)       &   17$^{\rm h}$ 40$^{\rm m}$ 44\fs589(4) \\
Declination (J2000)           &   $-$53$^\circ$ 40\arcmin 40\farcs9(1)  \\
Galactic longitude, $l$       &   338\fdg164                            \\
Galactic latitude, $b$        &   $-$11\fdg96                           \\
Dispersion Measure, DM (cm$^{-3}$pc)                  & 71.8(2)         \\
Period, $P$ (s)               &   0.00365032889720(1)                   \\
Period derivative, $\dot P$ (10$^{-20}$)              & 16.8(7)         \\
Period epoch (MJD)            &   51917.0                               \\
Orbital period (d)            &   1.35405939(5)                         \\
Projected semimajor axis (s)  &   1.65284(7)                            \\
Epoch of ascending node, T$_{\rm asc}$ (MJD)          & 51749.710822(6) \\
Eccentricity$^{\rm a}$, $e$   &   $<10^{-4}$                            \\
Data span (MJD)               &   51719$-$52116                         \\
Post-fit rms ($\mu$s)         &   114                                   \\
Mass function (M$_{\sun}$)    &   0.0026442(3)                          \\
Companion mass, $M_c$ (M$_{\sun}$)                    &   $>0.19$       \\ 
\enddata
\tablecomments{Numbers in parentheses represent three times the formal 
$1\sigma$ uncertainties given by {\sc tempo} in the least-significant 
digits reported.}
\tablenotetext{a}{All other parameters have been derived setting $e=0$ (see
text).}
\end{deluxetable}

\vskip -5truecm
\begin{deluxetable}{lll}
\scriptsize
\tablewidth{0.98\columnwidth}
\tablecaption{\label{tb2:tpar} {Parameters for possible companion}}
\tablecolumns{3}
\tablehead{\colhead{Parameter}&\colhead{He-WD}&\colhead{MS}}
\startdata
Companion mass, M$_{c}$ (M$_{\sun}$)        &  0.19         & 0.19--0.8  \\
Orbital inclination, {\it i} ($^\circ$)     &  90           &   90--17   \\
Orbital separation, $a$ (R$_{\sun}$)        &  6.0          &  6.0--6.7  \\
Roche lobe radius, R$_{L}$ (R$_{\sun}$)     &  1.3          &  1.3--2.2  \\
Eclipse radius, R$_{E}$ (R$_{\sun}$)        &  7.1          &  7.1--7.9  \\
Stellar radius, R (R$_{\sun}$)              &  $\sim 0.03$  &  0.2--0.96 \\
\enddata
\end{deluxetable}

\ifx\undefined\allcaps\def\allcaps#1{#1}\fi

\end{document}